\documentclass{aa}  

\usepackage[flushleft]{threeparttable}

\usepackage{graphicx}
\usepackage{txfonts}
\usepackage{lipsum}
\usepackage{subcaption}         
\usepackage{lscape}             
\usepackage{placeins}           
\usepackage{amsmath}	        
\usepackage{advdate}
\usepackage[]{hyperref}
\usepackage{xcolor}
\hypersetup{colorlinks,
            linkcolor=red,
            citecolor=blue,
            filecolor=cyan,
            urlcolor=magenta}
\usepackage{orcidlink}
\newcommand{\Msun}{\mbox{$M_{\odot}$}}
\newcommand{\AU}{\mbox{$\mathrm{au}$}}

\graphicspath{{./}{./}} 

\makeatletter
\let\linenumbers\relax
\makeatother

\begin{document}

   \title{Dust supply to close binary systems}


   \author{Francesco Marzari\inst{1}\orcidlink{0000-0003-0724-9987}
        \and 
           Gennaro D'Angelo\inst{2}\orcidlink{0000-0002-2064-0801}
        }

   \institute{Department of Physics and Astronomy, University of Padova, via Marzolo 8, I-35131, Padova, Italy\\
             \email{francesco.marzari@pd.infn.it}
            \and
              Theoretical Division, Los Alamos National Laboratory, Los Alamos, NM 87545, USA\\
             \email{gennaro@lanl.gov}
             }

   \date{Received \today}

 
  \abstract
   {
Binary systems can be born surrounded by circumbinary discs. The gaseous
discs surrounding either of the two stellar companions can have their life
extended by the supply of mass arriving from the circumbinary disc.
   }
   {
The objective of this study is to investigate the gravitational interactions 
exerted by a compact and eccentric binary system on the circumbinary and
circumprimary discs, and the resulting transport of gas and solids between
the disc components.
   } 
   {
We assume that the gas in the system behaves as a fluid and model its
evolution by means of high-resolution hydrodynamical simulations.
Dust grains are modeled as Lagrangian particles that interact with
the gas and the stars.
   }
   {
Models indicate that significant fluxes of gas and dust proceed from
the circumbinary disc toward the circumprimary disc. For the applied
system parameters, grains of certain sizes are segregated outside
the tidal gap generated by the stars.
Consequently, the size distribution of the transported dust is not
continuous but it presents a gap in the $\sim$mm size range.
In close binaries, the lifetime of an isolated circumprimary disc
is found to be short, $\sim 10^{5}$ years, because of its small mass.
However, because of the influx of gas from beyond the tidal gap,
the disc around the primary star can survive much longer, $\sim 10^{6}$ years,
as long as gas accretion from the circumbinary disc continues.
The supply of solids and the extended lifetime of a circumbinary
disc also aids in the possible formation of giant planets. 
Compared to close binary systems without a circumbinary disc,
we expect a higher frequency of single- or multiple-planet systems. 
Additionally, a planetesimal or debris belt can form in proximity of 
the truncation radius of the circumprimary disc and/or around the location
of the exterior edge of the tidal gap.
   }
   {}


   \titlerunning{Dust Supply to Close Binary Systems}
   \authorrunning{Marzari and D'Angelo}

   \maketitle

\section{Introduction}

About $50$\% of main sequence FGK stars have companions \citep[for a review, see][]{offner2023} 
with a separation distribution that peaks around $40\,\AU$. Data also show that 
the distribution of mass ratios is almost uniform. 
These features are expected to be the outcome of 
the formation process and of post-formation evolution. Most planets in binaries 
are found in S-type orbits, that is, they move around one of the two components 
of the binary. An updated catalog of exoplanets in binary star systems can be found 
at the web site\footnote{\url{https://adg.univie.ac.at/schwarz/multiple.html}}
of Richard Schwarz \citep{cata2016} at the University of Vienna.
So far, $217$ planets have been discovered in binary systems and some of them,
like HD~59686, HD~7449, $\gamma$~Cep, HD~4113, HD~41004, 30~Ari, Gliese~86, 
and HD~196885 \citep{su2021} have separations smaller than $30\,\AU$.
In the range between $10$ and $200\,\AU$ (close to intermediate binaries) 
the perturbations of the companion star are expected to significantly affect 
the planet formation process.  

Among the different mechanisms that can hinder planet formation in binaries, 
we can list 
\begin{itemize} 
\item  Gas heating due to the formation of spiral waves \citep{nelson2000}, 
which can prevent the condensation of ices and thus reduce the amount of solid 
material available for planet assembly;
\item  Increased disc eccentricity \citep{marzari_ecc1_2009,marza_ecc2_2012,kley_cephei_gam_2012},
which is expected to excite higher relative velocities among growing dust 
particles, and thus slow or prevent their growth;
\item Formation of hydraulic jumps \citep{picogna2013} at the sites of
spiral waves, which can induce vertical motion of the dust and reduce sedimentation 
towards the mid-plane of the disc;
\item Increased complexity of the dynamics of planetesimals, possibly slowing
the accumulation into larger bodies \citep{marz_hans2000,thebault2006};
\item Truncation of the circumstellar disc due to the gravitational perturbations
of the companion star \citep{lubov94}.
\end{itemize}

Of all these adverse effects, the disc truncation may be the most critical 
since it could significantly decreases the lifetime of the disc and reduce 
the amount of gas and dust available for the formation of planets, in particular 
the giant ones. An extreme case is that of  Gliese~86 b, which formed in a disc
that was probably truncated within a few $\AU$ from the primary star, 
by the gravitational perturbations of the companion \citep{zeng2022}. 
Nonetheless, all close to intermediate binaries with planets share a similar 
problem, related to the reduced reservoir of mass and possible short disc lifetime. 
For example, HD~59686 has a separation of $13.6\,\AU$ and an eccentricity 
of $0.73$ \citep{HD59686} while HD~196885 has a separation of $21\,\AU$ and 
eccentricity of $0.42$ \citep{HD196885}, parameters similar to another system 
harboring planets, such as the well-known $\gamma$~Cephei \citep{neuhauser2007}. 
In all these systems, the circumprimary disc would be truncated to small radial 
distances (a few \AU) from the primary star. 
 
A mechanism that can affect the mass of a circumprimary disc, and extend 
its lifetime, is the concurrent presence of a circumbinary disc. Mass can 
flow from it toward the inner disc(s), delaying their dispersal 
\citep{dutrey1994,moni2007,nelson2016}. Because of this feeding mechanism, 
both gas and solid material may be available for planet growth over 
timescales regulated by the circumbinary material. 

In this paper, we use numerical simulations to explore the possibility that 
both gas and dust can flow from a circumbinary disc toward the inner region
surrounding the the binary stars, supplying material to their discs. 
\cite{nelson2016} studied a similar scenario applied to the case of the GG~Tau
system, using SPH calculations and focusing on the gas flow. 
We achieve a significantly higher resolution, by using a nested grid code
\citep{gennaro2013}, in both the circumprimary and the circumbinary disc, 
and include dust evolution. We adopt a test bench for our experiments based
on a binary system similar to $\gamma$~Cephei, whose disc around the primary 
star has been extensively studied by means of hydrodynamics simulations 
\citep{kley_nelson2008,kley_cephei_gam_2012,lucas_gamma_ceph2021}. 
This system can be considered as a prototype of close binary systems in which
the binary companion is strongly affecting the circumprimary disc.

In Section~\ref{sec:CBS}, we provide further details on the motivations 
of this study. The methods and algorithms we use are described in 
Section~\ref{sec:MA}. Results on the gas and dust dynamics in the system,
including gas and solids accretion rates, are presented in
Section~\ref{sec:SD}. Some implications of the results regarding
planetary assembly around the primary star are discussed in Section~\ref{sec:IPF}.
Our conclusions are presented in Section~\ref{sec:Con}.

 \section{Close binary systems}\label{sec:CBS}

\cite{marz2019} found a peak in the distribution of binary systems hosting 
exoplanets with a separation of about $20\,\AU$ (see their Figure~3). 
This feature may be a bias introduced by the detection process but, 
if otherwise confirmed, 
it would suggest that planet formation can and does occur in an environment 
strongly perturbed by the gravitational tides of the close companion star. 
In Table~\ref{tab:close_binaries}, we list $8$ binary systems hosting exoplanets
with semi--major axis in the range between $19$ and $36\,\AU$ and high eccentricities,
ranging from $0.38$ to $0.5$. By far the most extensively studied system 
in this list is $\gamma$~Cephei, which has become a sort of standard model 
for testing planet formation scenarios in binaries. 
The evolution of the circumprimary disc, from which the planet may have formed, 
has been extensively modeled with hydrodynamics calculations
\citep[e.g.,][]{kley_nelson2008,paard2008, kley_cephei_gam_2012,lucas_gamma_ceph2021}.
These models show that the disc around the primary is truncated around 
$\approx 25$\% of the binary separation ($4$--$5\,\AU$) and it generally develops 
a significant eccentricity, which may affect both dust accumulation and planetsimal 
dynamics. 
\cite{marzari_ecc1_2009,marza_ecc2_2012} used hydrodynamics calculations to study
a somewhat wider configuration, with a binary semi--major axis of $30\,\AU$ and 
different values of the binary eccentricity, testing how these parameters
would affect the evolution of the disc.  

In all of the above-mentioned studies, it is manifest that the truncation radius
is small and the disc may lose mass relatively quickly (if not re-supplied),
rendering the process of planet formation quite intricate, at least according 
to classic core accretion models of formation developed for discs around single 
stars . 
Either the assembly of planets is a very robust process, which can occur also 
in some extreme conditions, or it follows somewhat different routes around binaries 
than it does around single stars.  
Therefore, it is important to study the individual steps of planet formation, 
starting from the early phases of evolution of the circumstellar disc, to figure out 
if there may be large differences compared to formation in a single-star scenario. 
For this reason, we focus here on the evolution of a circumprimary disc in a system 
similar to $\gamma$~Cephei, hypothesizing the existence of a (resolved) circumbinary 
disc. The main goal is to assess whether, and to what extent, 
the truncated circumbinary disc can supply gas and solids to the circumprimary disc. 

\begin{table}
	\centering
	\caption{Selected list of close binary systems hosting planets
        (from \url{https://exoplanet.eu/planets_binary/}).
        }
	\label{tab:close_binaries}
	\begin{tabular}{lcccc} 
		\hline
		System & $M_\mathrm{A}$ [\Msun] & $M_\mathrm{B}$ [\Msun] & $a$ [\AU] & $e$\\
		\hline
		$\gamma$~Cephei & $1.29$ & $0.384$ & $19.56$ & $0.410$\\
            HD~196885       & $1.10$ & $0.508$ & $19.78$ & $0.420$\\
            HD~41004~A      & $0.70$ & $0.400$ & $20.00$ & $0.400$\\
            30~Ari~B        & $1.22$ & $0.520$ & $21.90$ & $0.380$\\
            Gliese~86       & $0.87$ & $0.540$ & $23.70$ & $0.429$\\
            HD~7449         & $1.05$ & $0.170$ & $34.65$ & $0.301$\\
            HD~8673         & $1.35$ & $0.380$ & $35.00$ & $0.500$\\
            HD~126614       & $1.15$ & $0.320$ & $36.20$ & $0.500$\\
		\hline
	\end{tabular}
	    \begin{tablenotes}
      \small
      \item  The list
        includes name, component masses, semi-major axis and 

		   \hskip 0.1 truecm eccentricity.
    \end{tablenotes}
\end{table}

There are discrepancies in the literature regarding the masses of the
components of the $\gamma$~Cephei system. 
\citet{neuhauser2007} reported the values $M_\mathrm{A}=1.40 \pm 0.12\,\Msun$
and $M_\mathrm{B}=0.409 \pm 0.018\,\Msun$ whereas \citet{torres2007} estimated
the values $M_\mathrm{A}=1.18 \pm 0.11\,\Msun$ and
$M_\mathrm{B}=0.362 \pm 0.022\,\Msun$.
More recently, \citet{mugrauer2022} derived dynamical masses for the components 
of $M_\mathrm{A}=1.294 \pm 0.081\,\Msun$
and $M_\mathrm{B}=0.384 \pm 0.013\,\Msun$.
In our reference model, described more in details in the next section, 
we choose the masses provided by \citet{neuhauser2007}, resulting in
a mass ratio $M_\mathrm{B}/M_\mathrm{A}=0.286$. 
We point out that, in terms of gas dynamics, what matters are not the masses
of the stars but rather their ratio. In this respect, all three solutions referred
to above provide a ratio $\approx 0.3$.
The orbital semi-major axis is $a=20\,\AU$ and the eccentricity is $e=0.4$
\citep[][who also used the masses reported by \citeauthor{neuhauser2007}]{endl2011}.
These properties do not change over time.

\section{Methods and algorithms}\label{sec:MA}

The calculations presented herein assume that the binary stellar system 
is surrounded by a disc of gas and solids, referred to as circumbinary disc. 
The inner boundary of this disc is largely set by the tidal field of the stars, 
as they gravitationally interact with the disc material, and by viscous 
stresses within the gas.
Although the circumbinary disc can be truncated, mass may still flow toward 
the stars. We aim to quantify this flow, or lack thereof. Separate discs may 
also be present around one or both stars. 
This occurrence, however, is hindered by the eccentricity of the binary orbit, 
which limits the accumulation of gas, especially around the lower-mass secondary
\citep{gennaro2006}.
If the supply of material (gas and solids) from the circumbinary disc 
is negligible, due to the strong tidal field of the binary, these discs would 
basically evolve in isolation, possibly exchanging mass, a process that our
models can also quantify.

We assume a two-dimensional (2D)
geometry in which stars, gas and solids orbit in the same plane.
Local quantities are either vertically integrated (e.g., mass density)
or averaged (e.g., temperature, velocity).
Polar coordinates $\{r, \phi\}$, with origin on the primary star, are used
to describe the system.
The choice of the origin is arbitrary and does not affect the outcomes of
the calculations. When using a non-inertial reference frame, the algorithm
takes into account all non-inertial forces.
It should be noted that, although the gas in the farthest parts of the circumbinary
disc is expected to rotate about the center of mass of the binary system,
the circumprimary gas rotates about the primary star. 
The radial variable, $r$, has 
units of the semi-major axis of the binary, $a$. In the reference model,
the radial domain extends from $0.04\,a$ to $144\,a$. 
The azimuth angle varies between $0$ and $2\,\pi$. 
The large outer radius of the domain is chosen so that the circumbinary
disc can evolve freely, without interfering with artificial boundaries.
Indeed, the large grid allows us to model the precession of the entire
circumbinary disc about the binary's center of mass.

The coordinate system rotates around the primary star at a variable 
rate, which is imposed by requiring that the azimuth angle of the secondary
star is a constant and the angular velocity of the secondary is zero 
\citep[for details, see][and references therein]{gennaro2006}. 
The disc material evolves in the gravitational potential 
\begin{equation}
 \Phi= -\frac{GM_\mathrm{A}}{r}-\frac{GM_\mathrm{B}}%
             {\sqrt{|\mathbf{r}-\mathbf{r}_\mathrm{B}|^{2}+\varepsilon^{2}}}
       + \Phi_\mathrm{I},
 \label{eq:PHII}
\end{equation}
in which $M_\mathrm{A}$ and $M_\mathrm{B}$ are the masses of the primary 
and secondary stars, respectively, $\mathbf{r}$ is the position vector, 
$\mathbf{r}_\mathrm{B}$ is the position vector of the secondary, 
and $\Phi_\mathrm{I}$ is the potential arising from non-inertial forces. 
The quantity $\varepsilon$ is a constant introduced to avoid singularities
at the position of the secondary, and is set equal to a fraction (a few 
to several percent) of the Hill radius of the secondary,
$R_{\mathrm{H}}=r_{\mathrm{B}}\left[M_\mathrm{B}/(3M_\mathrm{A})\right]^{1/3}$.
Self-gravity of the disc material is ignored.
Note that since $R_{\mathrm{H}}$ is much larger than the pressure scale height,
$H$, at the secondary (see below), $\varepsilon$ in Equation~(\ref{eq:PHII}) 
should be related to the Hill radius rather than to $H$.

The gas in the system is treated as a compressible fluid that evolves according
to the Navier–Stokes equations, which are solved employing a finite-difference 
code that is second-order accurate in space and time 
\citep[see][and references therein]{gennaro2013}.
To reduce some restrictions on the time-step of the calculations, we apply 
an orbital advection algorithm as described in \citet{gennaro2012}.
In the base model, the disc is described by a grid with $9002\times 404$
cells in the radial and azimuthal direction, respectively.
Non-reflective boundary conditions \citep{gennaro2006} are applied at 
the outer radial boundary. Reflective or outflow boundary conditions
are applied at the inner radial boundary. Outflow boundary conditions
are used when the system has relaxed, to estimate the mass flux toward 
the primary star.

We use a local-isothermal equation of state for the gas, in which the
vertically-integrated pressure is $P \propto \Sigma T$ and the gas 
temperature is $T\propto r^{-3/7}$. This assumption results in a flared
disc with an aspect ratio $H/r\propto r^{2/7}$. 
At the distance of the secondary, this ratio is $0.04$.
The initial surface density of the gas is an axisymmetric power-law,
equal to $\Sigma\propto (a/r)^{3/2}$
and extending to the outer boundary of the computational domain.
After settling, the total amount of gas in the system is 
$\approx 0.02\,M_\mathrm{A}$.
We apply a kinematic viscosity $\nu\propto r^{3/2}$, so that
the initial conditions would allow for a steady-state circumbinary disc 
(i.e., $d(\nu\Sigma)/dr\approx 0$).
The kinematic viscosity at the distance of the secondary corresponds 
to a turbulent viscosity parameter $\alpha\approx 0.001$ \citep{S&S1973}.
To test the effects of this parameter, another simulation was executed 
with $\alpha\approx 0.005$.

In order to resolve the discs around the stars, together with the entire
circumbinary disc, the reference model uses a hierarchy of $3$ nested grids 
\citep[see][and references therein]{gennaro2013}. The basic grid represents
the full domain identified above. The sub-grids cover the same azimuthal 
domain as the basic grid, $2\,\pi$ around the primary, and have the same 
inner radial boundary, $r=0.04\,a$. 
The second grid level extends out to $r=72\,a$ whereas the third grid level
extend out to $r=4.8\,a$. The linear resolution, in both radial and azimuthal
direction, increases by a factor $2$ for each added grid level.
The outer radial boundaries of the sub-grids were chosen upon conducting
preliminary tests. The second grid level was designed to include the entire
circumbinary disc (which has an outer radius, as discussed below); the third
grid level was designed to include the region around the binary, hence its outer
boundary was placed far from the secondary but inside the edge of the tidal gap
(see Section~\ref{sec:SD}).

\subsection{Gas tracing}\label{sec:GT}

Gas trajectories are tracked by means of passive tracers (mass-less particles), 
which are advected by the fluid. Tracers allow for a precise determination 
of fluid paths, in both stationary and non-stationary flows.
The equations of motion of the tracers are integrated over the time-step
$\Delta t$ of the hydrodynamics solver by interpolating gas velocities 
at the locations of the tracers and by advancing their positions in time 
via a second-order Runge-Kutta method. Spatial and temporal interpolations 
are performed by using gas velocities on the most refined grid level
to which each tracer belongs. The spatial interpolation applies 
a monotonized harmonic mean \citep{vanleer1977}, which is second-order 
accurate and capable of handling discontinuities and shock conditions.
Therefore, as the hydrodynamics scheme, the trajectories are second-order 
accurate in both space and time.

Since solids with short stopping times (small Stokes numbers) are well-coupled
to the gas, tracers can also be used as a proxy to describe the motion of fine 
dust (Stokes numbers $\lesssim 0.01$). 
For the conditions applied in these models, exterior to the gap region
and between $r\approx 10$--$20\,a$, tracers can be used to approximate 
the motion of dust grains smaller than $10$--$100\,\mu$m. 
Because of the rapidly declining gas density, this size reduces to 
$\approx 1\,\mu$m around $r = 30\,a$.
In the circumprimary disc, however, tracers can describe the motion of larger 
particles, up to $\sim 1\,\mathrm{cm}$ in radius, because of the much higher 
gas density.

To identify regions from which gas and dust can possibly filter across the 
circumbinary gap and feed the circumprimary disc, tracers are deployed 
in contiguous annular regions of the computational domain, according to 
the radial distance from the primary.
The outermost region, $10<r<20\,a$, is mostly exterior to the gap
(which, however, is eccentric). Tracers are released in four additional
annular regions, down to $r=0.1\,a$. The innermost region, which extends 
between $0.1$ and $0.5\,a$, includes the outer part of the circumprimary 
disc. This region is used to track possible transfer of gas and dust 
between the circumprimary disc and the secondary star. 
A total number of $180\,000$ tracers is used in this study.

\subsection{Evolution of solids}\label{sec:ES}

The evolution of solids that are partially or weakly coupled to gas 
dynamics is modeled by means of a Lagrangean approach.
In this case, particles have mass and move according to applied
(gravity, drag and non-inertial) forces.
Details on the methods are provided in \citet[][and references therein]{gennaro2022}.
To evaluate the delivery of solid material to the circumprimary disc, 
we group solids in three size bins, containing particles of 
$1\,\mathrm{mm}$, $1\,\mathrm{cm}$ and $10\,\mathrm{cm}$ radius, $R_{s}$.
Together with the information provided by the gas tracers,
this approach allows us to quantify the supply of solid material,
from fine dust to pebble/cobble-size solids.
This piece of information represents a necessary condition 
for the formation of planets in the circumprimary disc.

We assume that the solids are made of silicates (quartz)
and simulate $12\,000$ particles in each size bin.
Their initial orbits are randomly distributed between $20\,a$ and $50\,a$ 
from the primary,
far away from the stars and well beyond the edge of the tidal gap.
At such distances, gas temperatures in the models are low, $\lesssim
20\,\mathrm{K}$. However, within the circumprimary disc, temperatures can
exceed $150\,\mathrm{K}$ \citep[see also][]{lucas_gamma_ceph2021} and,
therefore, solids smaller than $\approx 1\,\mathrm{cm}$ would rapidly
sublimate. For this reason we chose to model silicates instead of ices.

\section{System dynamics}\label{sec:SD}

\begin{figure*}
\centering%
\resizebox{1.8\columnwidth}{!}{\includegraphics[clip]{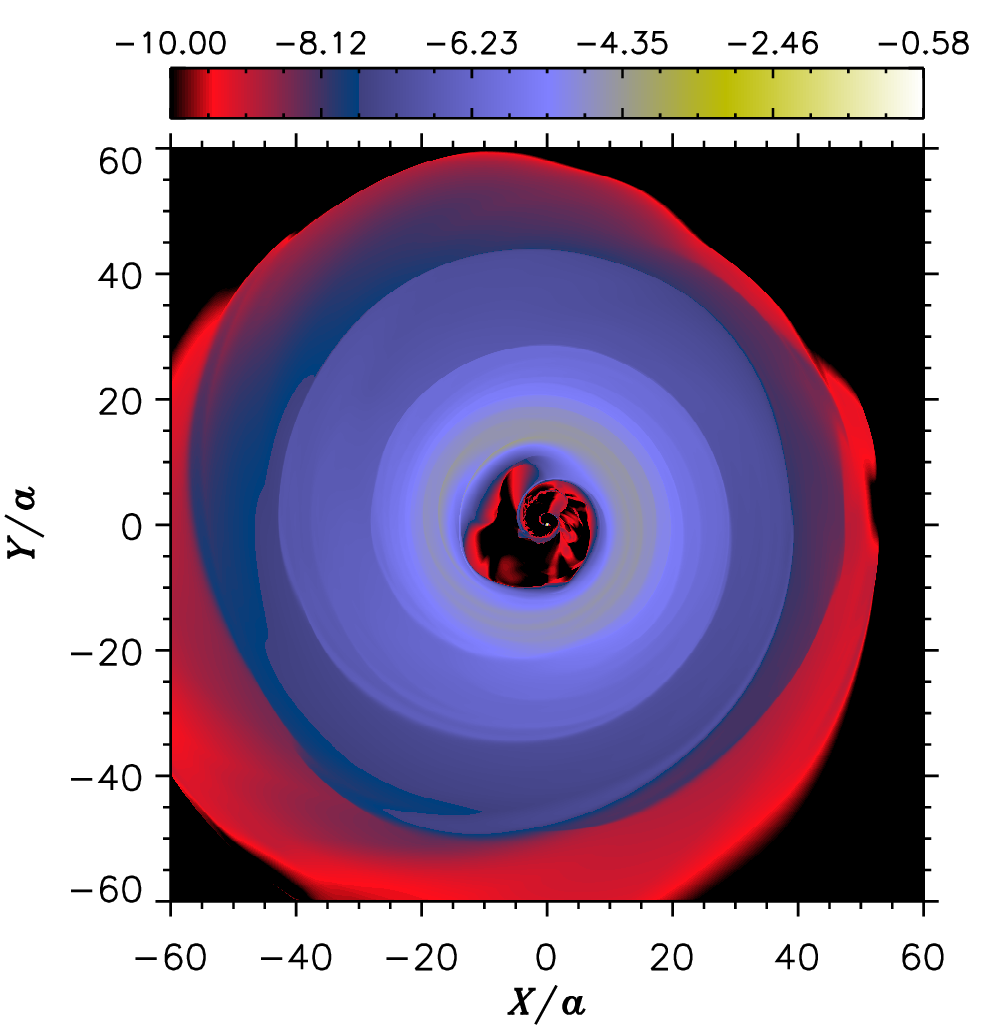}%
                             \includegraphics[clip]{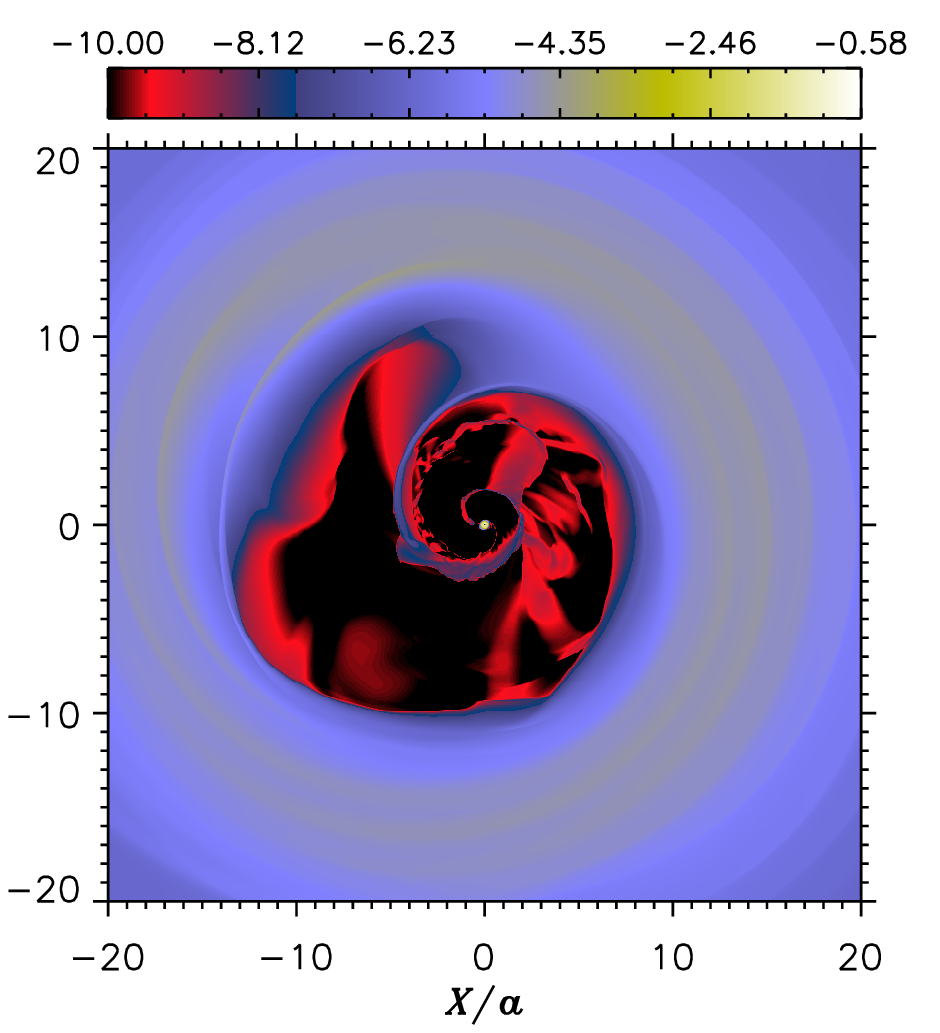}%
                             \includegraphics[clip]{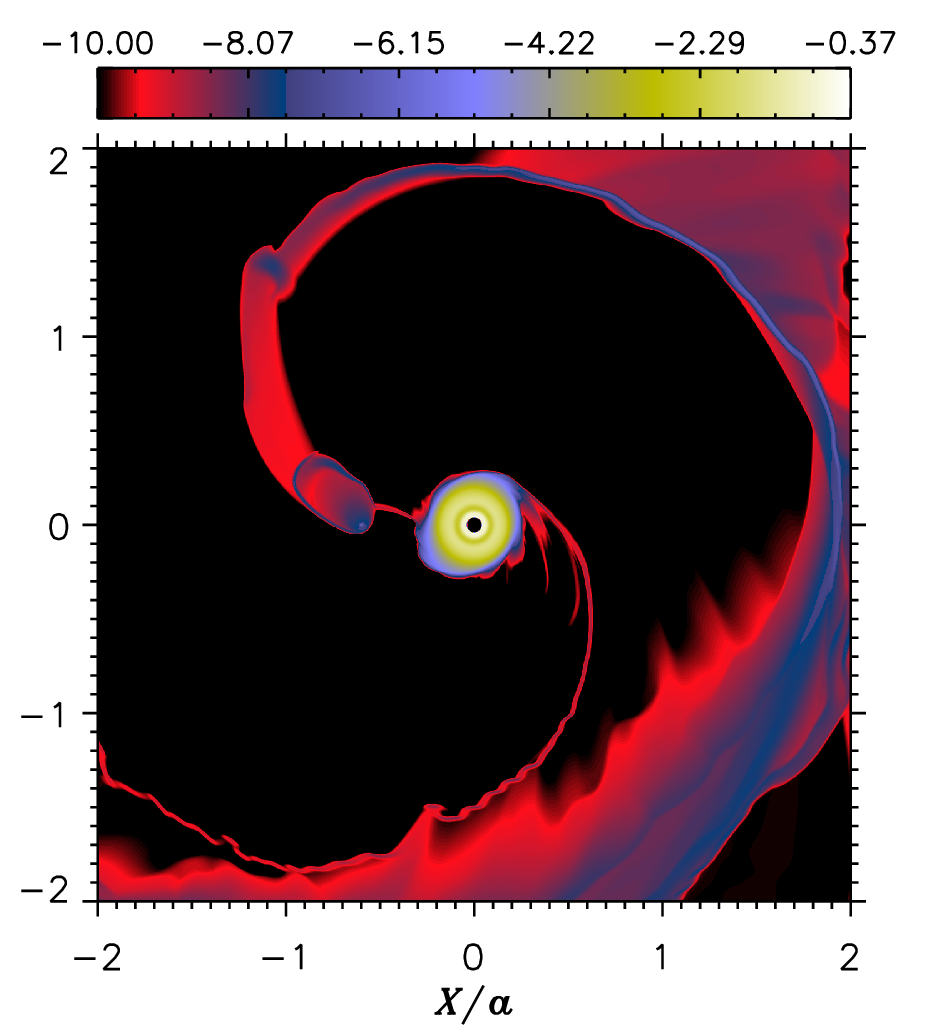}}
\resizebox{1.8\columnwidth}{!}{\includegraphics[clip]{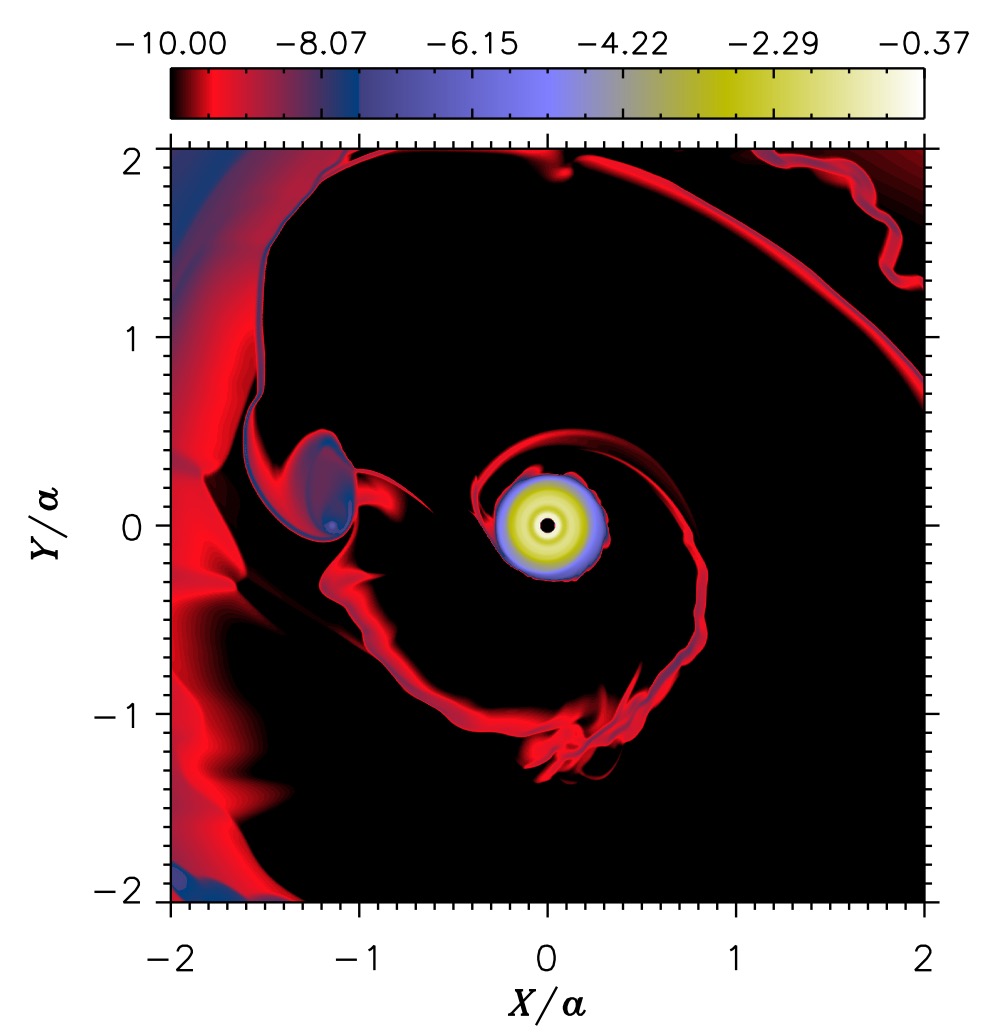}%
                             \includegraphics[clip]{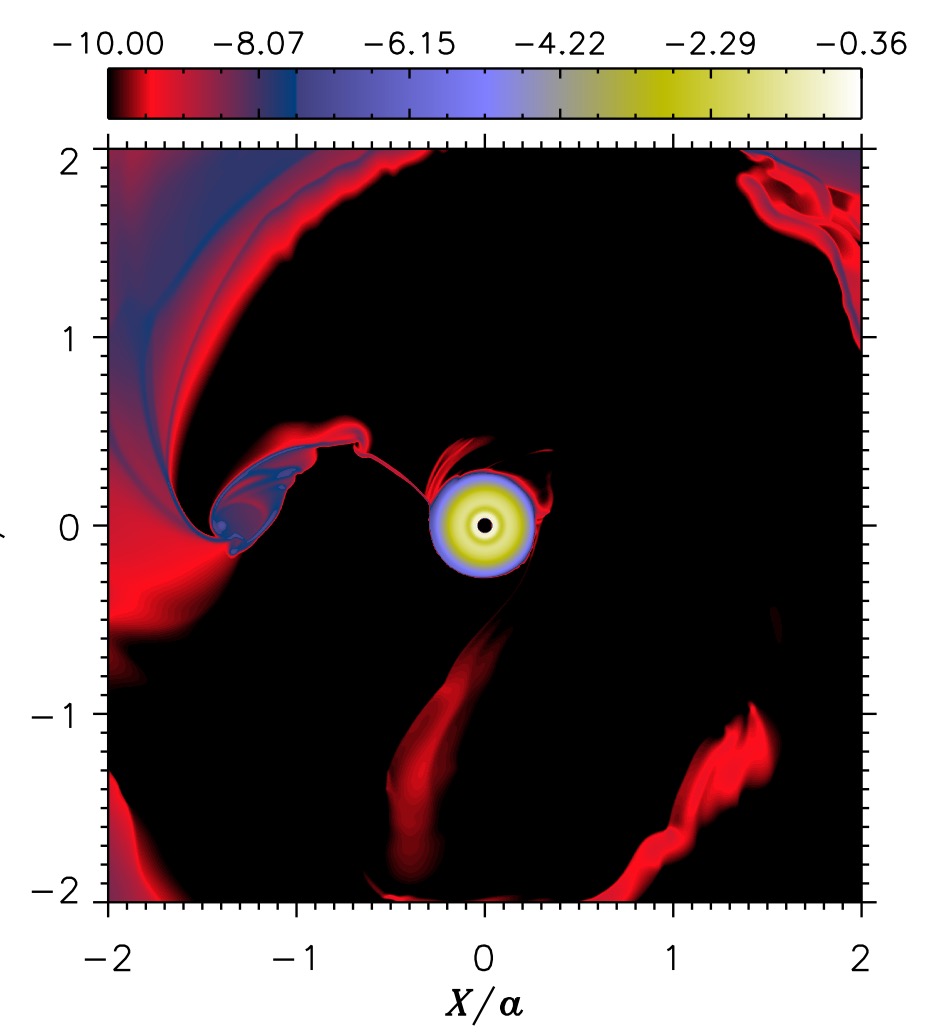}%
                             \includegraphics[clip]{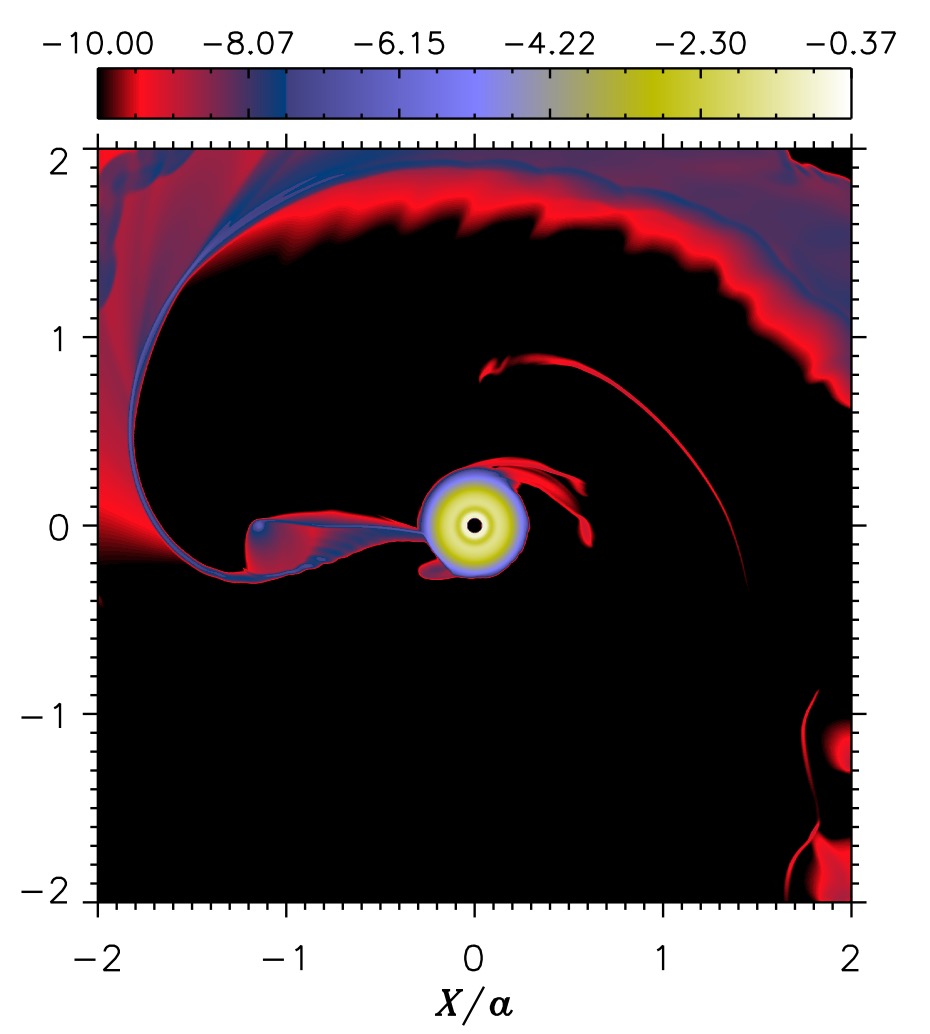}}
\caption{%
         Logarithm of the gas surface density in the base model,
         at time $t=3600\,T$.
         This model uses three overlapping polar grids, 
         the outermost of which reaches out to $144\,a$.
         In the top panels, the secondary star is located 
         at the pericenter of its orbit. The bottom panels
         show the gas surface density at different orbital phases, 
         with the secondary at apocenter in the central panel.
         The color bar represents $\Sigma$ in units of $M_\mathrm{A}/a^{2}$.
         For the adopted system parameters, a density of 
         $10^{-2}$ corresponds to $\approx 310\,\mathrm{g\,cm^{-2}}$.
        }
\label{fig:3610}
\end{figure*}

The gas surface density around the binary system is displayed
in Figure~\ref{fig:3610}, after $3600$ binary periods, $T$. 
The top-left panel includes the entire circumbinary disc, whose outer
radius appears to be around $60\,a$. This radius, however, may vary
according to definitions. Integrating the gas surface density over the region
$r\ge 9\,a$ (i.e., beyond the tidal gap), $99.9$\% of the total mass is contained 
within $r\approx 52\,a$.
Since the computational domain extends much farther, to $r=144\,a$, and the 
initial surface density is not truncated, this outer radius is not an artificial
effect of limited computational domain or initial conditions, but it is rather
determined by gas dynamics in the outer parts of the circumbinary disc.
The estimated volume density of the gas at $r\gtrsim 60\,a$ is
$\lesssim 10^{-20}\,\mathrm{g\,cm^{-3}}$.

The top-center panel of Figure~\ref{fig:3610} shows the gap region, 
whose outer edge has an azimuthally
average radius $r\approx 9\,a$. The gap edge, however, is not circular and
precesses about the center of mass of the system.
The region surrounding the two stars is shown in the top-right panel,
when the secondary star is at pericenter, and at other orbital phases
in the bottom panels.
The radius of the circumprimary disc too is ill-defined but, 
by integrating the gas surface density from the inner boundary of the grid 
($r=0.04\,a$) out to $r=5\,a$, that is, inside the gap, $99.9$\% of 
the total mass is contained within $r\approx 0.22\,a$
and $99.99$\% is contained within $r\approx 0.26\,a$. 

The surface density of the gas in the system is undetermined and our 
choices of initial conditions are obviously arbitrary. The application 
of a locally isothermal equation of state allows for hydrodynamical 
outcomes to be re-scaled, but the dynamics of the solids does depend 
on the initial condition.
Some upper bounds on the gas surface density can be estimated from 
the orbital evolution of the binary orbit driven by gravitational 
torques, under the assumption that the adopted orbit is close to
final. 
We performed such an experiment and allowed the orbit of the stars 
to change according to the torques exerted by the gas, and simulated
the evolving orbits for approximately $80$ binary periods.
Results indicate that the orbital eccentricity varies at an average rate
$\langle de/dt\rangle\approx 10^{-7}\,\mathrm{yr}^{-1}$, corresponding
to an evolution timescale $e/\dot{e}$ somewhat longer than the lifetime 
of the circumbinary disc.
The semi-major axis of the binary evolves much more slowly. 
During the simulated period, the average rate is 
$\langle da/dt \rangle\approx -3\times 10^{-8}\,a/\mathrm{yr}$, 
displaying $30$\% 
variations. The timescale $a/\dot{a}$ is therefore much longer than 
the lifetime of the circumbinary gas. Over the duration of the simulations
($\approx 3600\,T$), assuming these rates of change, the orbital eccentricity
would have changed by several percent whereas the semi-major axis would have
changed by much less.
Therefore, we can conclude that neglecting orbital variations in the models
is a reasonable approximation.

\subsection{Accretion rates of gas and fine dust}\label{sec:AR}

We estimated the accretion rates of gas toward the circumprimary
disc using tracer particles (see Section~\ref{sec:GT}), which provide
a description of the evolution of the gas, starting from the region 
of their deployment.
The transport can be quantified by using the conservation of the number
of the tracers. Indicating with $\mathcal{N}$ their number density,
conservation requires that
\begin{equation}
    \int\left(\frac{\partial{\mathcal{N}}}{\partial t}%
             +\nabla\cdot\mathbf{F}\right)dV =0,
\label{eq:div}
\end{equation}
where $\mathbf{F}=\mathcal{N}\mathbf{u}$ is the flux of tracers
($\mathbf{u}$ is the gas velocity). 
Equation~(\ref{eq:div}), integrated over the disc region 
identified by the radial boundaries $r_{1}$ and $r_{2}$ ($>r_{1}$), 
provides the accretion rate
\begin{equation}
    \frac{dM}{dt}=-\int\mathbf{F}\cdot\mathbf{\hat{n}}\,dS%
                 =\int \left(r_{2} F_{2}-r_{1} F_{1}\right)d\phi.
\label{eq:dmdt}
\end{equation}
Note that, since the geometry is two-dimensional, the integral in 
Equation~(\ref{eq:div}) is actually a surface integral, hence the
line integrals in Equation~(\ref{eq:dmdt}). Although $dM/dt$ represents 
a number per unit time (the domain integral of $\mathcal{N}$), 
it can be transformed into an accretion rate 
by converting the number density $\mathcal{N}$ into a surface density 
(e.g., grams per square centimeter). This is accomplished by normalizing 
the initial number of tracers, using the mass of the region in which 
they are released.
The interface fluxes, $F_{1}$ and $F_{2}$, are chosen as positive when  
directed toward the primary. 
In order to quantify average values of mass transfer, all quantities 
in Equation~(\ref{eq:dmdt}) are time-averaged over $\approx 100$ 
binary periods, $T$.

The two radii, $r_{1}$ and $r_{2}$, are chosen as representative of
the outer boundary of circumprimary disc ($0.5\,a$) and of the gap
($10\,a)$, respectively. 
The actual radius of the circumprimary disc is shorter (see above)
and the inner edge of the gap is not a circle. These details, however,
are not important. The region between $r_{1}$ and $r_{2}$ is strongly
perturbed by the tidal field of the stars, it has very low gas densities,
and thus it is expected to not represent a significant source of gas 
for the circumprimary disc.
If there is an influx of material toward the primary, it must come from
beyond the gap edge. For this reason, precise values for $r_{1}$ and
$r_{2}$ are unnecessary, as long as material flowing inside $r_{1}$ 
supplies the circumprimary disc and material beyond $r_{2}$ originates
from the circumbinary disc.
Equation~(\ref{eq:dmdt}) is used to derive the (time-averaged) integral 
of the inner flux, $\langle F_{1}\rangle$, from the other two quantities,
$\langle dM/dt\rangle$ and $\langle F_{2}\rangle$,
both of which are directly measured from the evolution of the tracers. 
As expected, tracer analysis indicates that
$|\langle dM/dt\rangle|\ll \int |\langle r_{2} F_{2}\rangle| d\phi$
(recall that $dM/dt$ here is the variation of mass in
the region between radii $r_{1}$ and $r_{2}$).
Therefore, from Equation~(\ref{eq:dmdt}), we have
\begin{equation}
    \int\langle r_{1} F_{1}\rangle d\phi \simeq \int\langle r_{2} F_{2}\rangle d\phi.
\label{eq:rF}
\end{equation}
Since the local-isothermal approximation allows for mass re-scaling,
indicating with $\Sigma_{\mathrm{ref}}$ the average density between 
$10\,a$ and $20\,a$, in units of $M_\mathrm{A}/a^{2}$, the average accretion 
rate of gas toward the circumprimary disc, estimated from the tracer 
analysis, is $\approx 7\times 10^{-2}\,\Sigma_{\mathrm{ref}}\,(M_\mathrm{A}/T)$.
For the values applied herein, the accretion rate would be
$\approx 2\times 10^{-8}\,M_{\sun}\,\mathrm{yr}^{-1}$.

The ratio of the mass available beyond the gap edge to the accretion
rate toward the circumprimary disc can be used to estimate the lifetime
of the circumbinary disc. Note that this ratio is insensitive to 
the choice of the initial surface density, or $\Sigma_{\mathrm{ref}}$,
because both terms scale as $\Sigma_{\mathrm{ref}}$, although it would 
depend on the applied gas viscosity and temperature (i.e., pressure
scale height). 
This ratio, approximately $1.1\,\mathrm{Myr}$, can be identified as 
an $e$-folding time for the evolution of the disc mass. Therefore, 
the estimate of the circumbinary disc lifetime would be 
around $3\,\mathrm{Myr}$. 
This value, however, neglects other possible means of gas removal, 
such as energetic radiation from the stars (disc photo-evaporation).

The analysis of the tracers released inside $0.5\,a$ indicates that 
there is no gas flowing out of the circumprimary disc, and hence
no transfer of mass toward the secondary. This result also implies
that the circumprimary disc can lose mass only via accretion toward
the primary. A simple estimate of the inward transport in 
the circumprimary disc, based on steady-state accretion disc theory
\citep{lynden-bell1974,pringle1981}, can be done by approximating
the accretion rate to $3\pi\Sigma\nu$ (averaged over the disc), 
which is on average roughly twice as large as the influx of gas from 
the circumbinary disc. 
A more accurate determination can be obtained by measuring the variation 
of the circumprimary disc mass during the same period of time over which 
tracers are evolved.
This estimate, which is very well defined because of the sharp decline
of the gas surface density beyond $r\approx 0.22\,a$, basically confirms
that the disc is depleting (and the primary star is accreting) at a rate, 
$\approx 2.3\times 10^{-6}\,(M_\mathrm{A}/T)$, or
$\approx 0.17\,\Sigma_{\mathrm{ref}}\,(M_\mathrm{A}/T)$,
about two times as large as the supply rate from the circumbinary disc.
At this rate, neglecting again gas removal by stellar irradiation
and assuming there was no supply of material from the circumbinary
disc, the circumprimary disc would last only a few times 
$10^{5}$ years, a length of time probably too short to allow for 
the formation of giant planets.
This outcome mostly depends on the small size of the disc and its limited
mass reservoir.
Therefore, the delivery of material from the circumbinary disc appears
to be critical in this context. 
In fact, because of external supply, the circumprimary gas would deplete 
until attaining a surface density such that the average product $3\pi\Sigma\nu$ 
becomes comparable to the gas accretion rate provided by 
the circumbinary disc, which would ultimately dictate also the stellar 
accretion rate. 
In the setup used for the base model, this equilibrium would entail
a surface density around the primary a factor of $\approx 2$ smaller
(than that in Figure~\ref{fig:3610})
and an actual lifetime of the circumprimary disc similar to that 
of the circumbinary disc, a few Myr, which may be enough to form 
giant planets \citep[e.g.,][and references therein]{stevenson2022}.

\begin{figure}
\centering%
\resizebox{1\columnwidth}{!}{\includegraphics[clip]{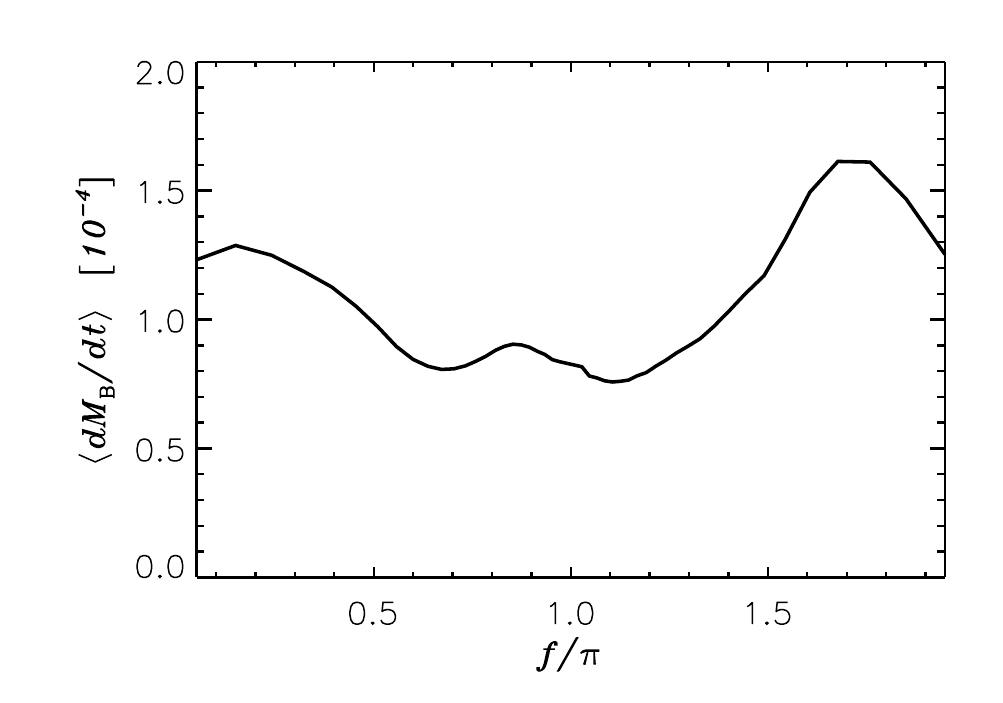}}
\caption{%
         Average accretion rate of gas on the secondary star as a function
         of the true anomaly, $f$, along its orbit. The results refer
         to the base model. In the plot, the accretion rate is averaged 
         over $100$ orbital periods of the binary.
         The pericenter is at $f=\pi$. The units are 
         $\Sigma_{\mathrm{ref}}\,(M_\mathrm{A}/T)$, where
         $\Sigma_{\mathrm{ref}}$ is the average density between 
         $10$ and $20\,a$, in units of $M_\mathrm{A}/a^{2}$.
        }
\label{fig:dmbdt}
\end{figure}
As mentioned above, the secondary star is not fed gas by the circumprimary
disc. It is however supplied by the gas inflow within the gap region. 
The fraction of this flux intercepted by the secondary is expected 
to be small and can be estimated directly from the simulations 
\citep[see, e.g.,][]{gennaro2006,gennaro2012}. This accretion rate is 
almost three orders of magnitudes lower than the accretion rate from
the circumbinary disc toward the circumprimary disc. 
In terms of the reference density introduced above, the rate would be 
$\approx 10^{-4}\,\Sigma_{\mathrm{ref}}\,(M_\mathrm{A}/T)$.
Figure~\ref{fig:dmbdt} shows the average accretion rate of gas on the
secondary, as a function of the true anomaly (see Figure's caption for
details). The accretion rate is modulated, with maximum delivery of gas
prior to apocenter passage \citep[see, e.g.,][]{pawel1996}.

As displayed in the top-right and bottom panels of Figure~\ref{fig:3610},
there is no permanent disc formation around the secondary, and the region
has relatively low density.
The average gas mass within $0.6$ Hill radii of the secondary star is
$\approx 2\times 10^{-4}\,\Sigma_{\mathrm{ref}}\,M_{\mathrm{A}}$, which
would correspond to $\approx 10^{-3}$ Earth's masses for the
values adopted in the model.

\subsection{Accretion rates of solids}\label{sec:ARS}

\begin{figure*}
\centering%
\resizebox{1.8\columnwidth}{!}{\includegraphics[clip]{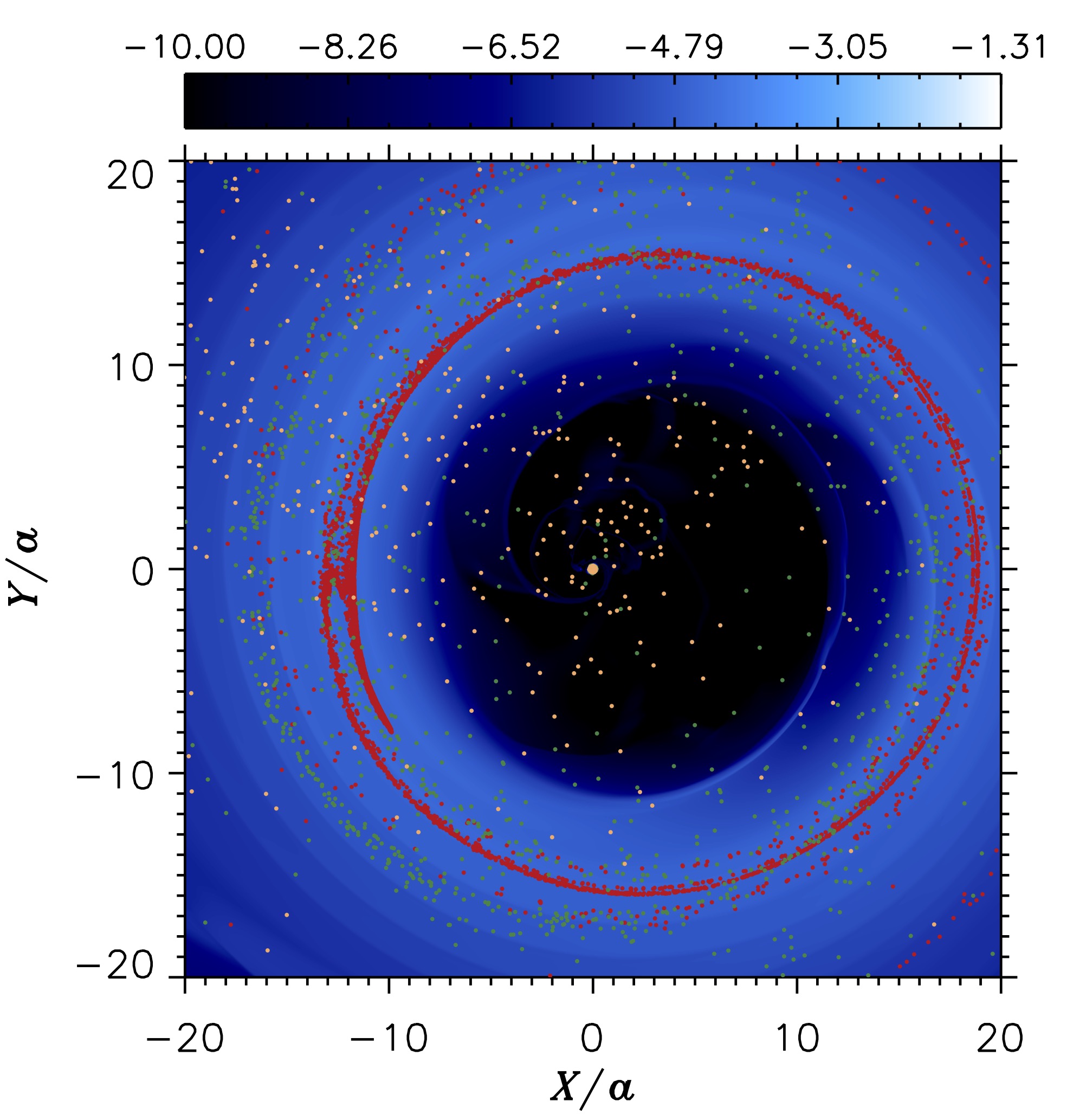}%
                             \includegraphics[clip]{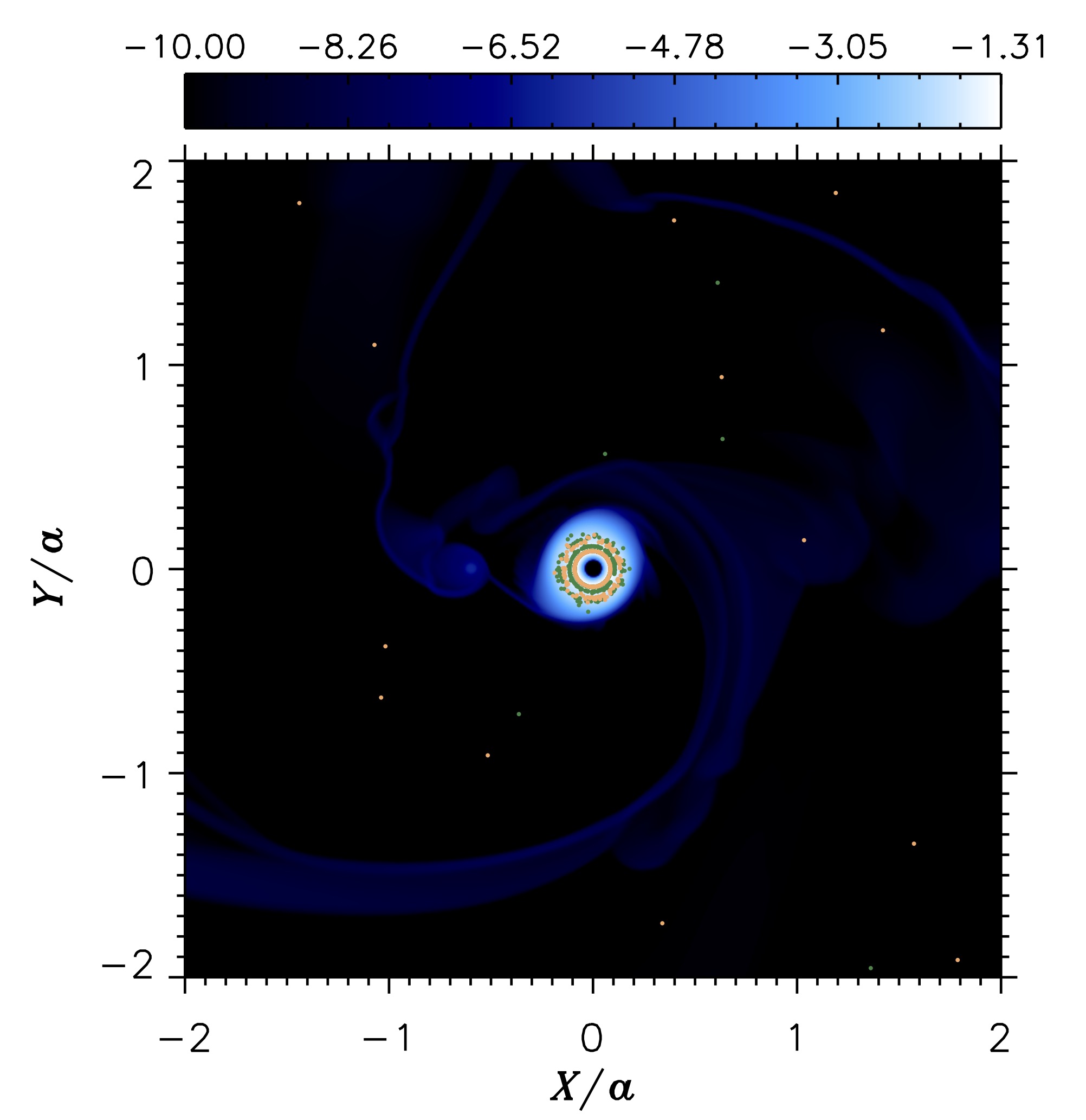}}
\caption{%
         Logarithm of the gas surface density in the base model with particles'
         positions superimposed: $1\,\mathrm{mm}$ (red),
         $1\,\mathrm{cm}$ (green), $10\,\mathrm{cm}$ (orange). All particles 
         are initially located beyond $20\,a$. 
         The panels show the gap region, its surroundings (left),
         and the circumprimary disc (right). The smallest particles 
         do not cross the gap region to reach the circumprimary disc
         whereas the larger particles do. 
         The secondary star is at apocenter.
         The color bar represents $\Sigma$ in units of $M_\mathrm{A}/a^{2}$.
        }
\label{fig:gpp}
\end{figure*}
The distribution of particles released in the circumbinary disc, superimposed
to the surface density of the gas, is plotted in Figure~\ref{fig:gpp}.
The particles are color-coded by their radius, $R_{s}$, as explained in 
the Figure's caption.
The smallest, $R_{s}=1\,\mathrm{mm}$, solids (red dots) remain confined beyond
the gap edge and do not move inward, across the gap region (see left panel).
Instead, the larger $1$ and $10\,\mathrm{cm}$ particles can be transferred from 
the circumbinary to the circumprimary disc, as indicated by the green and 
orange dots in the right panel. 

The segregation of the smallest particles outside the gap edge is a 
drag-driven effect, caused by the positive pressure gradient that develops
at those radial locations. The local rotation rate of the gas exceeds,
on average, that of the solids, generating a tail wind that tends to drive
the particles outward. Beyond the radial region of the positive pressure
gradient, particles tend to drift inward because gas rotation mainly
generates a head wind on them.
Therefore, the gap edge can represent a region of radial convergence
for some particles, despite the complex dynamics of the gas.
The converging effect is maximum for particles with a Stokes number of 
around unity, which are only partially coupled to the gas. At $r\approx 10\,a$, 
the thermodynamic conditions of the gas are such that $1\,\mathrm{mm}$ particles
have Stokes numbers close to $1$ and, therefore, are prone to be segregated
outside the tidal gap edge.
Since the Stokes number is inversely proportional to gas density, as 
the circumprimary disc depletes and densities reduce, $1\,\mathrm{mm}$ particles
around the gap edge are expected to transition toward an uncoupled regime and,
therefore, at some point during the evolution they may be released and move 
toward the circumprimary disc.

The Stokes numbers of $1$ and $10\,\mathrm{cm}$ solids are $\gg 1$, hence
they are less affected by drag at those locations. Their dynamics is mostly
dictated by the stars' gravity. In fact, they can acquire significant orbital 
eccentricity while orbiting in the circumbinary disc, which at some point
allows them to cross the gap region and to be captured (via drag) by the
high-density gas of the circumprimary disc.

\begin{figure}
\centering%
\resizebox{1\columnwidth}{!}{\includegraphics[clip]{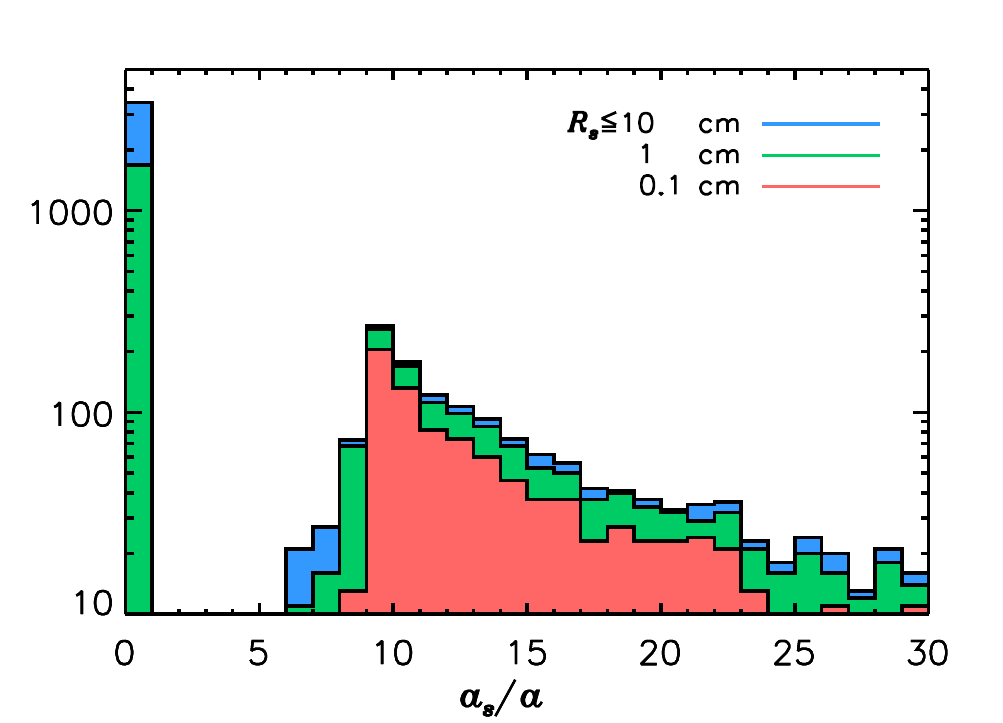}}
\caption{%
         Histogram of the semi-major axis of particles with
         different radii, as indicated in the legend. The histograms are stacked
         in order of increasing size: 
         the red histogram includes particles $R_{s}=1\,\mathrm{mm}$,
         the green histogram includes particles $R_{s}\le 1\,\mathrm{cm}$,
         and the blue histogram includes all particles.
         All solids start from beyond 
         $20\,a$ and only the larger particles drift toward
         the circumprimary disc. Results shown refer to the base model.
        }
\label{fig:a_hist}
\end{figure}

\begin{figure}
\centering%
\resizebox{1\columnwidth}{!}{\includegraphics[clip]{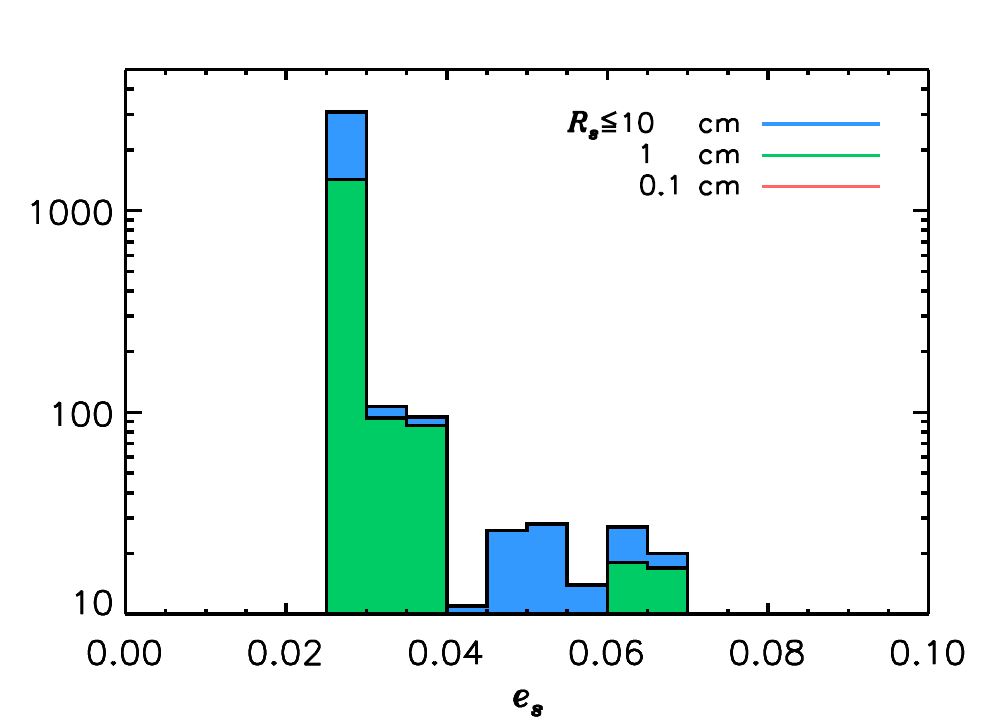}}
\caption{%
         Histogram of the orbital eccentricity of particles with
         different radii, as indicated.
         The histograms are stacked in order of 
         increasing size, as in Figure~\ref{fig:a_hist}.
         Only particles with semi-major axis smaller than $0.5\,a$ 
         (i.e., orbiting inside the circumprimary disc)
         are represented. The smallest, $1\,\mathrm{mm}$ particles 
         do not drift across the gap region, as they are held back 
         by the positive pressure gradient at the gap edge.
         Results refer to the base model.
        }
\label{fig:e_hist}
\end{figure}

Figure~\ref{fig:a_hist} shows the histograms of the semi-major axis of the particles
in Figure~\ref{fig:gpp}, which allows for a more quantitative assessment of the transfer
of solids toward the primary star and the pile-up of $1\,\mathrm{mm}$ dust around the
gap edge. (Recall that the solid particles are initially deployed beyond $20\,a$ 
from the primary.)
The histograms are stacked in order of increasing size: 
red for $R_{s}=1\,\mathrm{mm}$, green for $R_{s}\le 1\,\mathrm{cm}$ particles,
and for all particles.
In Figure~\ref{fig:a_hist}, the numbers of $1\,\mathrm{cm}$ and $10\,\mathrm{cm}$ 
particles transferred to the circumprimary disc are similar. Nonetheless, the
number of the $10\,\mathrm{cm}$ particles in the circumbinary disc (i.e., beyond
the gap edge) is smaller, because they are more efficiently ejected from the system,
indicating that their supply rate to the circumprimary disc is higher. 
In fact, during the last $50$ orbits of the binary, the $10\,\mathrm{cm}$ solids
are transferred toward the circumprimary disc at a rate about $1.8$ times that 
of the smaller, $1\,\mathrm{cm}$ particles.

The particles that arrive in the circumprimary disc settle on low-eccentricity
orbits after capture, as indicated by the histogram of orbital eccentricities
in Figure~\ref{fig:e_hist}. It is likely (because of the low density in 
the gap region) that particles are injected in the
circumprimary disc on highly-elliptical or even non-Keplerian orbits and, 
therefore, arrive at velocities much larger than those of resident solids.
These large relative velocities are likely to produce high-velocity collisions
among particles, leading to fragmentation into smaller dust or coagulation into
larger solids.
We do not model collisions. However, the results in Figure~\ref{fig:e_hist} 
indicate that, if particles avoid collisions, their orbital eccentricity is 
damped by gas drag over relatively short timescales. 
In fact, at the gas densities encountered in the
circumprimary disc, $1$--$10\,\mathrm{cm}$ solids have Stokes number $\ll 1$
and thus their dynamics quickly becomes well coupled to that of the gas.
The implication that, soon after upon arrival, these particles become 
basically indistinguishable from older, resident population.

The solid particles are evolved for $\approx 200\,T$ in the base model.
The supply rate of $1$ and $10\,\mathrm{cm}$ particles from beyond the gap 
region, toward the primary, initially grows over time but it later becomes 
relatively steady. We measured the average (over the last $\approx 100\,T$
of the solids' evolution) supply rate of the $1$ and $10\,\mathrm{cm}$ 
particles, from the variations of their number densities within
the circumprimary disc.
Once converted into accretion rates, they can be written as 
$\approx 80\,\sigma_{\mathrm{ref}}\,M_{\oplus}/T$,
for the $1\,\mathrm{cm}$ particles, and
$\approx 380\,\sigma_{\mathrm{ref}}\,M_{\oplus}/T$,
for the $10\,\mathrm{cm}$ particles. In these estimates, the quantity 
$\sigma_{\mathrm{ref}}$ represents the average surface density of solids, 
of the specific size, in the region between $10\,a$ and $20\,a$, 
in cgs units. 
For comparison,
the accretion rate of fine dust, estimated as a by-product of the accretion
rate of gas derived in Section~\ref{sec:AR}, can also be written as
$\approx \sigma_{\mathrm{ref}}\,M_{\oplus}/T$ (after converting the units
of the reference gas density $\Sigma_{\mathrm{ref}}$).

For the circumbinary disc setup applied herein, the often-quoted gas-to-solid 
mass ratio of $100$ would yield 
$\sigma_{\mathrm{ref}}\approx 4\times 10^{-3}\,\mathrm{g\,cm}^{-2}$.
Strictly speaking, however, this density would apply to fine dust whose dynamics 
is well-coupled to gas dynamics, resulting in an accretion rate
$\sim 6\times 10^{-5}\,M_{\oplus}\,\mathrm{yr}^{-1}$. 
Since the dynamics of larger solids is uncoupled from
that of the gas to a great extent, the estimate of $\sigma_{\mathrm{ref}}$
should also account for the removal rate of solids from the region around
the gap edge toward farther regions of the disc and for the supply rate
from those more distant regions. Therefore, a meaningful estimate of 
$\sigma_{\mathrm{ref}}$ for $1$ and $10\,\mathrm{cm}$ particles is 
more complex.
If we simply assume a value of $\sim 10^{-3}\,\mathrm{g\,cm}^{-2}$, 
the accretion rates above become 
$\sim 10^{-3}\,M_{\oplus}\,\mathrm{yr}^{-1}$ ($R_{s}=1\,\mathrm{cm}$)
and 
$\sim 5\times 10^{-3}\,M_{\oplus}\,\mathrm{yr}^{-1}$ ($R_{s}=10\,\mathrm{cm}$).
Note, however, that all these estimates require that the entire reservoir of solids
beyond the gap edge be available in the form of fine dust, $1$, or $10\,\mathrm{cm}$
particles. If instead the available mass was partitioned in the various size bins,
as is expected to be, the accretion rates in a given size bin would be lowered
accordingly. 

According to the estimated accretion rates and regardless of 
$\sigma_{\mathrm{ref}}$, the depletion time of $\approx 1$ to 
$\approx 10\,\mathrm{cm}$ particles, from the circumbinary disc between $10$ 
and $20\,a$, would be relatively short: 
from several hundreds binary periods, at the smaller end of 
the range, to around one hundred binary periods, at the larger end. 
If these timescales are too short compared to re-supply timescales of 
the solids from larger distances and from local coagulation/fragmentation, 
the flux of solids toward the primary may be also limited by these re-supply 
processes.
The smaller, $\sim 1\,\mathrm{mm}$ grains that remain trapped beyond 
the gap edge may also contribute to re-supply solids if they aid in 
the formation of larger particles.

Assuming a reference surface density 
$\sigma_{\mathrm{ref}}\sim 10^{-3}\,\mathrm{g\,cm}^{-2}$, equipartition 
of mass between $1$ and $10\,\mathrm{cm}$ particles, and inefficient 
re-supply of solids to the $10$--$20\,a$ disc region, about $30\,M_{\oplus}$ 
worth of solids would be transferred to the circumprimary disc
from beyond the gap region. 
This mass should be added to that already available locally.
If we estimate the local reservoir of solids in the circumprimary disc
by re-scaling the average surface density of the gas (say, by a factor
$100$), a few Earth masses of solids are available between
$2$ and $3\,\AU$ from the primary star 
(with the parameters applied herein). 

\subsection{High-viscosity model}\label{sec:HvM}

\begin{figure}
\centering%
\resizebox{1\columnwidth}{!}{\includegraphics[clip]{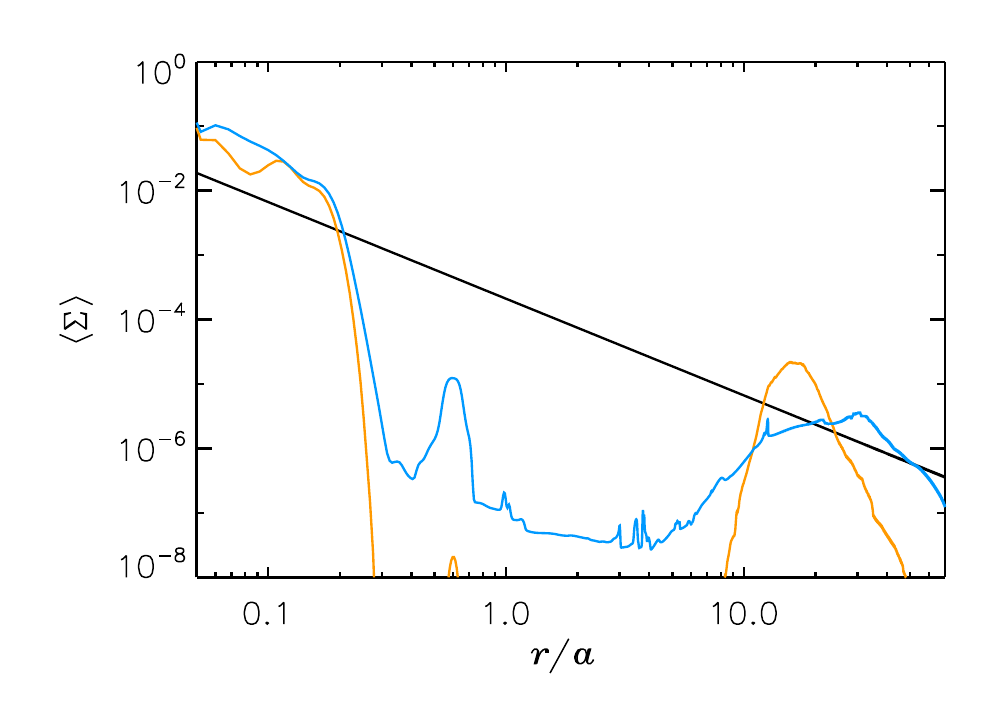}}
\caption{%
         Gas surface density averaged in azimuth around the primary,
         as a function of the radial distance, for the base model
         ($\alpha\approx 0.001$, orange curve) and a model with higher
         viscosity parameter, $\alpha\approx 0.005$ (light-blue curve).
         The Black line represents the initial condition.
         The units of $\langle\Sigma\rangle$ are $M_\mathrm{A}/a^{2}$.
         For the adopted system parameters, a density of 
         $10^{-2}$ corresponds to $\approx 310\,\mathrm{g\,cm^{-2}}$.
         The secondary star is locates at pericenter.
        }
\label{fig:sig_hvm}
\end{figure}

Figure~\ref{fig:sig_hvm} shows a comparison between the average surface
density of the gas around the primary star in the the base model
(orange curve) and another model with a higher turbulence parameter
(see Section~\ref{sec:MA}), $\alpha\approx 0.005$ (light-blue curve).
The two models are compared after an evolution of about $3300$ binary periods.
Mass and size of the circumprimary discs are similar in the two calculations.
Because of the larger viscous stresses that oppose tidal forces exerted by
the stars, the gap edge of the circumbinary disc extends inward to about 
$r\approx 5\,a$, closer to the stars than in the base model.
The densities in the circumbinary disc of the high-viscosity model are lower,
but the disc extends farther out than it does in the base model.
At large distances, the surface density also declines less abruptly than
it does in the base model.
Integrating $\Sigma$ form $r=5\,a$ outward, over $95$\% 
of the total mass of the circumbinary disc is contained within 
$r\approx 100\,a$. However, the mass of the circumbinary disc is very 
similar in the two models.
\begin{figure*}
\centering%
\resizebox{1.8\columnwidth}{!}{\includegraphics[clip]{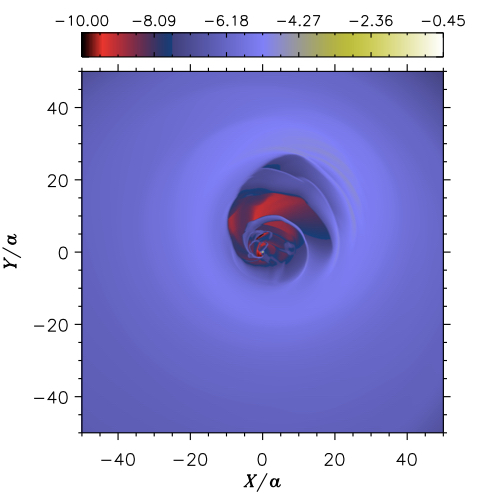}%
                             \includegraphics[clip]{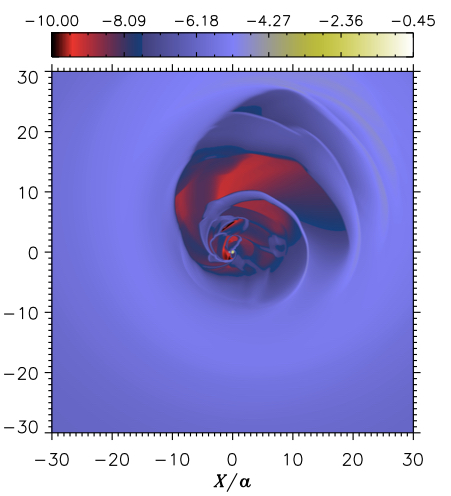}%
                             \includegraphics[clip]{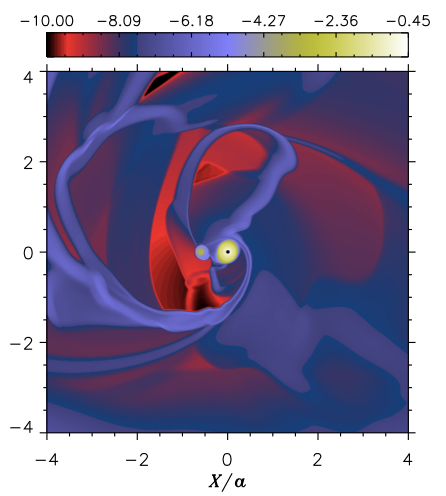}}
\caption{%
         Logarithm of the gas surface density in the high-viscosity model,
         at time $t=3300\,T$. As for the base model, the simulation uses
         overlapping polar grids that extend out to $144\,a$.
         The left and center panel show the circumbinary disc,
         whereas the right panel shows the circumprimary disc.
         The secondary star is located at the pericenter of its orbit.
         The color bar represents $\Sigma$ in units of $M_\mathrm{A}/a^{2}$.
         For the adopted system parameters, a density of 
         $10^{-2}$ corresponds to $\approx 310\,\mathrm{g\,cm^{-2}}$.
        }
\label{fig:3320}
\end{figure*}

The 2D density distribution around the stars is illustrated in Figure~\ref{fig:3320}.
The inner part of the circumbinary disc is shown in the left panel. 
Compared to Figure~\ref{fig:3610}, the perturbations induced by 
the density waves propagating beyond the tidal gap are less prominent,
as they are more efficiently damped by gas viscosity. As can be seen in
the center panel, the gap edge is highly asymmetrical around the binary,
significantly more than in the base case (compare with top-center panel
in Figure~\ref{fig:3610}).
In fact, the smooth decline of $\Sigma$ inside the gap 
(light-blue curve in Figure~\ref{fig:sig_hvm}) is mainly the result
of the azimuthal average, around the primary, of the asymmetrical
gas distribution within the gap region.

The secondary star is located at pericenter in both Figures~\ref{fig:sig_hvm}
and \ref{fig:3320} (visible in the right panel),
in correspondence of the build-up of gas in the gap at $r=0.6\,a$.
Peak gas densities around the secondary are about $30$ times as large as
in the surrounding gap regions. Although not fully visible in Figure~\ref{fig:sig_hvm},
similar density ratios are obtained around the secondary in the base model.
Nonetheless, as indicated in the figure, the gas around the secondary is
much denser than in the base model, by $2.5$--$3$ orders of magnitude,
an outcome mainly dictated by the much shallower gap of the high-viscosity model. 

\section{Implications for planet formation}\label{sec:IPF}

In single star systems, the dust density within the circumstellar gas 
is expected to decline with time, at all dust sizes, due to grain growth and 
dust transport processes \citep{testi2013}. Grain growth is a progressive 
process whereby the smaller (initially $\mu$m--size) dust particles accumulate 
into larger, pebble-size solids, which then can accumulate into planetesimals 
and possibly planets. Drag forces due to the interaction between grains 
and gas lead to a radial drift with a velocity relative to the gas that 
depends on the grain size and gas properties. The particles spiral towards 
the star on different timescales, although at some locations they may be 
halted by positive pressure gradients in the gas (due to, e.g., gas density 
enhancements generated by the gravity of a planet). 
However, the rapid decay of the dust mass in circumstellar gas, expected 
on the basis of grain growth and radial drift, is not measured in the surveys 
of $1$--$3\,\mathrm{Myr}$-old circumstellar discs observed by ALMA
\citep{testi2013}. It has been suggested \citep{bar2022,turrini2019} that 
the growth of giant planets causes the formation of second-generation dust 
through collisions of left-over planetesimals, whose orbits are excited 
by the presence of planets. 
Under these circumstances, the dust density decline would start only 
$1$--$2\,\mathrm{Myr}$ after the formation of giant planets. 

While in the case of single stars the dust density resurgence would just
involve part of the primordial reservoir of solids (i.e., those formed out 
of the dust present in the circumstellar gas around the star since formation),
in the binary star case with a circumprimary disc fed by an extended 
circumbinary disc, new $\sim\mu$m--size dust can be supplied during 
the lifetime of the circumbinary disc. Additionally, 
under appropriate conditions, a non--continuous size distribution of particles,
presenting a gap in some size range (mm-sizes, in our base model), 
would also 
be supplied to the circumprimary disc. These larger particles would presumably 
form out of coagulation/fragmentation exterior to the gas gap of the binary.
The mass partition between $\sim\mu$m--size dust and $1$--$10\,\mathrm{cm}$
particles depends on the assumptions made and on the details of the 
coagulation/fragmentation processes.
Segregation of particles beyond the gap edge is determined by the local
pressure scale height, density, and viscosity of the gas.
The range of particle sizes prevented
from crossing the gap would depend on the combinations of these parameters.
As these property of the gas evolve, segregated particles may be released 
from the gap edge region and move toward the circumprimary disc.

As mentioned at the end of Section~\ref{sec:ARS}, the circumbinary disc may
contain a primordial reservoir of a few Earth masses worth of solids. 
If these solids turned relatively quickly and efficiently into planetesimals, 
which then assembled into a planet a few times $M_{\oplus}$, the external supply 
of cm-size particles could aid in the formation of a larger (possibly giant) 
planet. 
We estimated a total influx of solid mass, from the circumbinary disc, 
of around $30\,M_{\oplus}$ in the form of $1$--$10\,\mathrm{cm}$ particles.
At an accretion efficiency of $10$\% \citep{gennaro2024}, approximately
$3\,M_{\oplus}$ would be added to a forming planet. Once attained several 
Earth masses, gas accretion may be significant, which could lead to growth
beyond the pebble isolation mass of the planet \citep{morby2012,bi2018},
halting any further inward drift of these particles. Therefore, even 
if the external supply of $1$--$10\,\mathrm{cm}$ particles was short-lived 
(see Section~\ref{sec:ARS}), significant amounts of pebble-size solids 
could accumulate in the outer part of the circumprimary disc. This
situation could possibly lead to enhanced planetesimal formation,
e.g., via streaming instability \citep{joha2015} or through turbulent gas 
concentration \citep{heng2010,hart2020}. The latter mechanism may be 
particularly effective in the binary case since the companion's 
perturbations excite strong spiral waves, which can contribute 
to turbulence generation in the gas. 
In the region beyond the planet's orbit (out to the truncation radius,
if the system is compact as the one under study), a higher planetesimal 
formation rate may be expected, compared to the single star case,
because of the steady supply of solids from the circumbinary disc. 
The resulting ring of planetesimals may trigger additional planet formation
or, otherwise, lead to the production of a dense debris disc in the system,
in the proximity of the truncation radius of the circumprimary disc.
This structure would be a significant observable feature, which may point
to a past influx of solids from larger distances and, therefore, to the past
presence of a circumbinary disc. This feature could also point to the possible
existence of planetary bodies inside the debris belt.

On a larger scale, if small grains remain trapped beyond the edge of the tidal
gap until the circumbinary gas eventually disappears, they too may be
observable as a debris disc. However, whether or not the original population 
of trapped particles would be preserved, depends on coagulation and fragmentation
processes. The enhanced density of grains that are collected around those locations,
as dust moves inward from farther regions of the circumbinary disc, can
ultimately produce a much broader size distribution of grains via collisions.

As mentioned above, solids arriving from the circumbinary to the circumprimary
disc may have a non-continuous size distribution.
This outcome is primarily determined by the local pressure gradient around
the gap edge region and by the Stokes numbers of the solids.
For example, the base model predicts a gap in the $\sim$mm size range,
because Stokes numbers are around unity in that range. 
It is not clear how such a size 
distribution of grains would affect the turbulent accumulation of dust 
into planetesimals.
It is possible that the resulting size distribution of planetesimals is different
from that generated by a \textit{continuous} distribution of dust grains, as 
it is likely to occur in discs around single stars.
It is difficult to argue whether or not these differences may actually discriminate
between single star and compact binary systems. In general, the cores of
giant planets can efficiently form out of planetesimals from a few to hundred 
$\mathrm{km}$ in radius. As long as the reservoir of planetesimals 
include these bodies, formation would occur similarly in both types of systems.
Probably, such an outcome would be favored by conditions in the circumbinary disc
that allowed a more continuous size distribution of dust to move through the gap.

More extreme cases are, however, conceivable. If viscous stresses were much
weaker than in the base model (e.g., smaller $\alpha$), both gas and dust could
be prevented from moving inward. Not only the availability of solids around
the primary star would be severely limited but also the gas would dissipate quickly
(in a few time $10^{5}$ years, see Section~\ref{sec:AR}).
It is unlikely that giant planets would be able to form under these conditions.

An intermediate situation can be envisaged, in which the core of a giant planet
would be able to form in the circumprimary disc, initiated by the local solids, 
and then aided by the supply of both gas and dust from the circumbinary disc.
However, if the circumbinary disc was short-lived, say one $\mathrm{Myr}$ or so,
the circumprimary disc would be short-lived as well, preventing the cores
to fully develop into gas giants. As a result,  gas-poor (or not-so-gas-rich)
planets may form, such as Uranus and Neptune.

\section{Conclusions}\label{sec:Con}

We simulated the evolution of the dust and gas in the environment surrounding
a compact binary system, in which a circumprimary disc co-exists with a 
large circumbinary disc.  
To model the evolution of the system, we used a high-resolution Eulerian code
that employs a nested-grid technique to resolve different spatial scales, both
much smaller and much larger than the stellar separation. In addition, tracers
were used to better track gas dynamics, as it streams from the outer circumbinary
disc to the inner circumprimary disc. A Lagrangian approach was used to simulate 
the evolution of the dust.

According to model results, significant differences should arise when comparing
the system evolution to that of a compact binary that is not supplied with gas
and dust by a massive-enough circumbinary disc. There may also be differences
with discs around single stars.

Assuming an average surface density of the circumbinary disc gas, between $10\,a$
and $20\,a$, of $\bar{\Sigma}$, in $\mathrm{g\,cm^{-2}}$, the accretion rate
on the circumprimary disc is 
$\approx 5\times 10^{-8}\,\bar{\Sigma}\,M_{\sun}\,\mathrm{yr}^{-1}$.
This estimate is based on a model with a viscosity parameter $\alpha\approx 0.001$
and a flared disc whose $H/r$ is $0.04$ at a distance $a$ from the primary star
(see Section~\ref{sec:MA}). 
The adopted setup, $\bar{\Sigma}\approx 0.4\,\mathrm{g\,cm^{-2}}$, results 
in an accretion rate of 
$\approx 2\times 10^{-8}\,\bar{\Sigma}\,M_{\sun}\,\mathrm{yr}^{-1}$.
Without such supply of gas, the lifetime of the circumprimary disc would be
limited to a few times $10^5$ years because of its small mass.
The supply of gas from beyond the gap region extends this timescale to about
$3\,\mathrm{Myr}$, the lifetime of the circumbinary disc.

Dust grains would also be transferred from large distances to the circumprimary
disc. Small, $\mu$m-size grains would be carried by the accreting gas in which
they are entrained whereas larger, $1$--$10\,\mathrm{cm}$-size grains would be 
driven across the gap mainly by the gravity of the stars.
The applied conditions of temperature and viscosity cause $\sim\mathrm{mm}$-size 
grains to be segregated beyond the gap edge because of the positive and large 
pressure gradient generated by the tidal field of the stars. This feature in
the gas is conducive to segregation of solids with order-of-unity Stokes numbers
due to gas-dust interactions.

As mentioned above,
since the dynamics of the larger particles is mostly governed by gravity forces,
they tend to acquire significant orbital eccentricity, which brings them toward
the primary star. Solids delivered to the circumprimary disc on highly elliptical
orbits may have spent long enough time in the cold circumbinary gas to collect
volatile materials (by condensation). Since their journey toward the circumprimary
disc is rapid, they can carry, and deliver, significant amounts of volatile 
material to the circumprimary gas.

This continuing influx of gas and solids is likely to have relevant implications
for the formation of giant planets. It is generally difficult to conceive that
such planets may be able to form within a few $10^5$ years (the lifetime of the
isolated circumprimary disc). The extended lifetime ($>1\,\mathrm{Myr}$) would 
allow a core of several Earth masses to acquire a significant, or massive, envelope.
The implication is that giant planets in compact binary systems are expected 
to be much more frequent in those systems with a significant circumbinary disc. 
The base model would allow for about $30\,M_{\oplus}$ worth of solids to be transferred
to the circumprimary disc, contributing to the accumulation of a giant planet core
and to the formation of planetesimals.

The size distribution of the solids flowing across the gap is likely to depend
on the density, temperature and viscosity of the inner parts of the circumbinary disc. 
This distribution may be non-continuous, as that resulting from the base model.
The ultimate effects on planetesimal formation of a non-continuous size
distribution are difficult to predict. As long as planetesimals of a few to 
hundred $\mathrm{km}$ in radius are available, gas giant cores can be assembled.
If a planet in the circumprimary disc exceeds its ``pebble-isolation'' mass,
the continuing influx of solids may lead to the formation of additional 
planets, or to the accumulation of a relatively massive debris or planetesimal belt
in the proximity of the truncation radius of the circumprimary disc. 
This potentially observable feature would also represent a strong indication
that the binary system was born with a circumbinary disc. 
Beyond the edge of the tidal gap, trapped grains may also form a debris belt,
independent of the inner one,
possibly providing further observational evidence of the past existence of
a long-lived circumbinary disc.

\begin{acknowledgements}
The authors would like to thank an anonymous reviewer, whose constructive
comments and suggestions helped improve the quality of this paper, and
the Editor for helpful feedback. 
G.D.\ acknowledges support from NASA’s Research Opportunities in Space 
and Earth Science.
Computational resources supporting this work were provided by 
the NASA High-End Computing (HEC) Program through the NASA Advanced 
Supercomputing (NAS) Division at Ames Research Center.
\end{acknowledgements}

%
\bibliographystyle{aa} 
\bibliography{example} 
\end{document}